\documentstyle[aps,epsf,rotate,preprint]{revtex}

\begin{document}

\draft

\title{Stochastic Feedback and the Regulation of Biological Rhythms}

\author{Plamen~Ch.~Ivanov$^1$,~Lu\'{\i}s~A.~Nunes~Amaral$^{1,2}$,
Ary~L.~Goldberger$^3$,~and~H.~Eugene~Stanley$^1$}

\address{$^1$ Center for Polymer Studies and Department of Physics,
  		Boston University, Boston, MA 02215\\
	$^2$ Condensed Matter Theory, Physics Department, Massachusetts
		Institute of Technology, Cambridge, MA 02139 \\ 
	$^3$ Cardiovascular Division, Harvard Medical School, 
		Beth Israel Hospital, Boston, MA 02215 }

\maketitle
\begin{abstract}
We propose a general approach to the question of how biological
rhythms spontaneously self-regulate, based on the concept of
``stochastic feedback''.  We illustrate this approach by considering
the neuroautonomic regulation of the heart rate. The model generates
complex dynamics and successfully accounts for key characteristics of
cardiac variability, including the $1/f$ power spectrum, the
functional form and scaling of the distribution of variations, and
correlations in the Fourier phases. Our results suggest that in
healthy systems the control mechanisms operate to drive the system
away from extreme values while not allowing it to settle down to a
constant output.
\end{abstract}

The principle of homeostasis asserts that biological 
systems seek to maintain a constant output after perturbation
\cite{Bernard}. Recent evidence, however, indicates that healthy
systems even at rest display highly irregular dynamics
\cite{Akselrod81}.  Here, we address the paradox of how to reconcile
homeostatic control and complex variability.  The concept of dynamic
equilibrium or homeostasis \cite{Bernard} led
to the proposal that physiological variables, such as the cardiac
interbeat interval $\tau(n)$, maintain an approximately constant value
in spite of continual perturbations. Thus one can write in general
\begin{equation}
\tau(n) = \tau_0 + \eta\,,
\label{e-rate0}
\end{equation}
where $\tau_0$ is the ``preferred level'' for the interbeat interval
and $\eta$ is a white noise with strength $\sigma$, defined as the
standard deviation of $\eta$.

\begin{figure}
\narrowtext
\centerline{
\epsfysize=1.4\columnwidth{\epsfbox{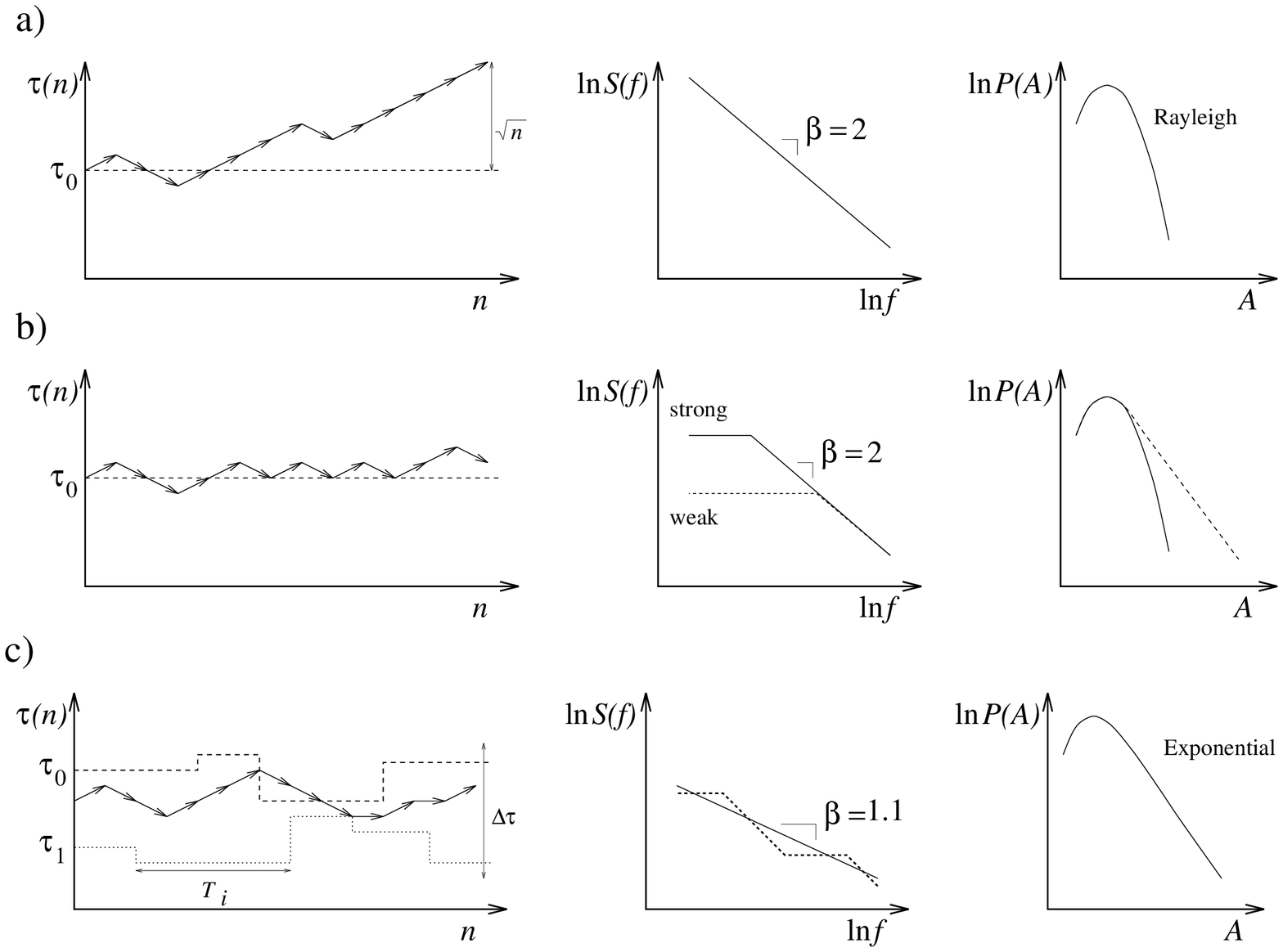}}}
\vspace*{-3.5cm}
\caption{ Schematic representation of the dynamics of the model. (a)
Evolution of a random walk starting from initial position $\tau_0$.
The deviation of the walk from level $\tau_0$ increases as $n^{1/2}$,
where $n$ is the number of steps. The power spectrum of the random
walk scales as $1/f^2$ (Brownian noise).  The distribution $P(A)$ of
the amplitudes $A$ of the variations in the interbeat intervals
follows a Rayleigh distribution.  (b) Random walk with a bias toward
$\tau_0$.  For short time scales (high frequencies), the power
spectrum scales as $1/f^2$ (Brownian noise) with a crossover to white
noise at longer time scales due to the attraction to level
$\tau_0$. Note the shift of the crossover to longer time scales (lower
frequencies) when stronger noise is present.  However, in both cases,
$P(A)$ follows the Rayleigh distribution.  (c) Random walk with two
stochastic feedback controls. In contrast to (b), the levels of
attraction $\tau_0$ and $\tau_1$ change values in time.  Each level
persists for a time interval $T_i$ drawn from a distribution with an
average value $T_{\rm lock}$.  Perturbed by changing external stimuli,
the system nevertheless remains within the bounds defined by
$\Delta\tau$ even after many steps.  We find that such dynamical
mechanism, based on a single characteristic time scale $T_{\rm lock}$,
generates a $1/f$ power spectrum over several decades.  Moreover,
$P(A)$ decays exponentially, which we attribute to nonlinear Fourier
phase interactions in the walk.}
\label{f-walk}
\end{figure}

We first re-state this problem in the language of random walks.  The
time evolution of an uncorrelated and unbiased random walk is
expressed by the equation $\tau(n+1) - \tau(n)= \eta$. The deviation
from the initial level increases as $n^{1/2}$, so an uncorrelated and
unbiased random walk does not preserve homeostasis
(Fig.~\ref{f-walk}a).  To maintain a constant level, there must be a
bias in the random walk \cite{Wax54},
\begin{mathletters}
\begin{equation}
\tau(n+1) - \tau(n) = I(n)\,,
\end{equation}
with
\begin{equation}
I(n)  = \cases{ w~ \left( 1 + \eta \right) \,, & if $\tau(n) 
			< \tau_0$, \cr
               -w~ \left( 1 + \eta \right) \,, & if $\tau(n) 
			\ge \tau_0$. \cr }
\end{equation}
\label{e-feedback}
\end{mathletters}
The weight $w$ is the strength of the feedback input biasing the walker
to return to its preferred level $\tau_0$. 

We find that Eq.~(\ref{e-feedback}) does not reproduce the statistical
properties of the empirical data (Fig.~\ref{f-walk}b).  We therefore
generalize Eq.~(\ref{e-feedback}) to include several inputs $I_k$
($k=0,1,\cdots,m$), with different preferred levels $\tau_k$, which compete
in biasing the walker, and Eq.~(\ref{e-feedback}a) becomes
\begin{mathletters}
\begin{equation}
\tau(n+1) - \tau(n) = \sum_{k = 0}^{m} I_k(n)\,,
\end{equation}
where
\begin{equation}
I_k(n)  = \cases{ w_k~ \left( 1 + \eta \right) \,, & if $\tau(n) 
			< \tau_k$, \cr
               -w_k~ \left( 1 + \eta \right) \,, & if $\tau(n) 
			\ge \tau_k$. \cr }
\end{equation}
\label{e-input}
\end{mathletters}

From a biological point of view, it is clear that the
preferred levels $\tau_k$ of the inputs $I_k$ cannot remain constant in
time, for otherwise the system would not be able to respond to varying
external stimuli. We assume that each preferred interval $\tau_k$ is a
random function of time, with values correlated over a time scale
$T_{\rm lock}^k$.  We next coarse grain the system and choose
$\tau_k(n)$ to be a random step-like function constrained to have values
within a certain interval and with the length of the steps drawn from a
distribution with an average value $T_{\rm lock}$ (Fig.~\ref{f-walk}c).
This model yields several interesting features
not fully explained by traditional models: (1)
$1/f$ power spectrum, (2) stable scaling form for the distribution of
the variations in the beat-to-beat intervals and (3) Fourier phase
correlations \cite{Mackey77}.

To illustrate the approach for the
specific example of neuroautonomic control of cardiac dynamics, we first
note that the healthy heart rate is determined by three major inputs:
(i) the sinoatrial (SA) node; (ii) the parasympathetic (PS); and (iii)
the sympathetic (SS) branches of the autonomic nervous system.

(i) The SA node or pacemaker is responsible 
for the initiation of each heart beat;  in the absence of
other external stimuli, it is able to maintain a constant interbeat
interval \cite{Bernard}.  Experiments in which PS and SS inputs are
blocked reveal that the interbeat intervals are very regular and average
only 0.6s \cite{Berne96}. 
The input from the SA node, $I_{SA}$, thus
biases the interbeat interval $\tau$ toward its intrinsic level
$\tau_{SA}$ (see Fig.~\ref{f-walk}b).

(ii) The PS fibers conduct impulses that slow the heart rate. 
Suppression of SS stimuli, while under PS regulation, can result in
the increase of the interbeat interval to as much as 1.5s
\cite{Berne96}. The activity of the PS system changes with
external stimuli. We model these features of the PS input, $I_{PS}$, by
the following conditions: (1) a preferred interval, $\tau_{PS}(n)$,
randomly chosen from an uniform distribution with an average value {\it
larger\/} than $\tau_{SA}$, and (2) a correlation time, $T_{PS}$,
during which $\tau_{PS}$ does not change, where $T_{PS}$ is drawn from a
distribution with an average value $T_{\rm lock}$.

(iii) The SS fibers conduct impulses that speed up the heart 
beat.  Abolition of parasympathetic influences when the sympathetic
system remains active can decrease the interbeat intervals to less
than 0.3s \cite{Berne96}. There are several centers of sympathetic
activity highly sensitive to environmental influences
\cite{Berne96}. We represent each of the $N$ sympathetic inputs by
$I_{SS}^j$ ($j=1,\cdots,N$).  We attribute to $I_{SS}^j$ the following
characteristics: (1) a preferred interbeat interval
$\tau_{SS}^j(n)$ randomly chosen from a uniform distribution with an
average value smaller than $\tau_{SA}$, and (2) a correlation time
$T_j$ in which $\tau_{SS}^j(n)$ does not change;  $T_j$ is drawn from
a distribution with an average value $T_{\rm lock}$ which is the same
for all $N$ inputs (and the same as for the PS system), so $T_{\rm
lock}$ is the {\it characteristic time scale} of both the PS and SS
inputs.  

  The characteristics for the PS and SS inputs correspond to a random
walk with stochastic feedback control (Fig.~\ref{f-walk}c).  Thus, for
the present example of cardiac neuroautonomic control, we have $N+2$ inputs
and Eq.~(\ref{e-input}a) becomes
\begin{equation}
\tau(n+1) - \tau(n)  = I_{SA}(n) + I_{PS} \left( n, \tau_{PS}(n) 
		\right) + \sum_{j = 1}^{N} I_{SS}^j
		\left( n, \tau_{SS}^j(n) \right)\,,
\label{e-heartinput}
\end{equation}
where the structure of each input is identical to the one in
Eq.~(\ref{e-input}b). Equation~(\ref{e-heartinput}) cannot fully
reflect the complexity of the human cardiac system. However, it
provides a general framework that can easily be extended to include
other physiological systems (such as breathing, baroreflex control,
different locking times for the inputs of the SS and PS systems
\cite{Akselrod81,Berne96}, etc.). Equation~(\ref{e-heartinput})
captures the essential ingredients responsible for a number of
important statistical properties of the healthy heart rate.

\begin{figure}
\narrowtext
\vspace*{0.0cm}
\centerline{
\epsfysize=1.5\columnwidth{\epsfbox{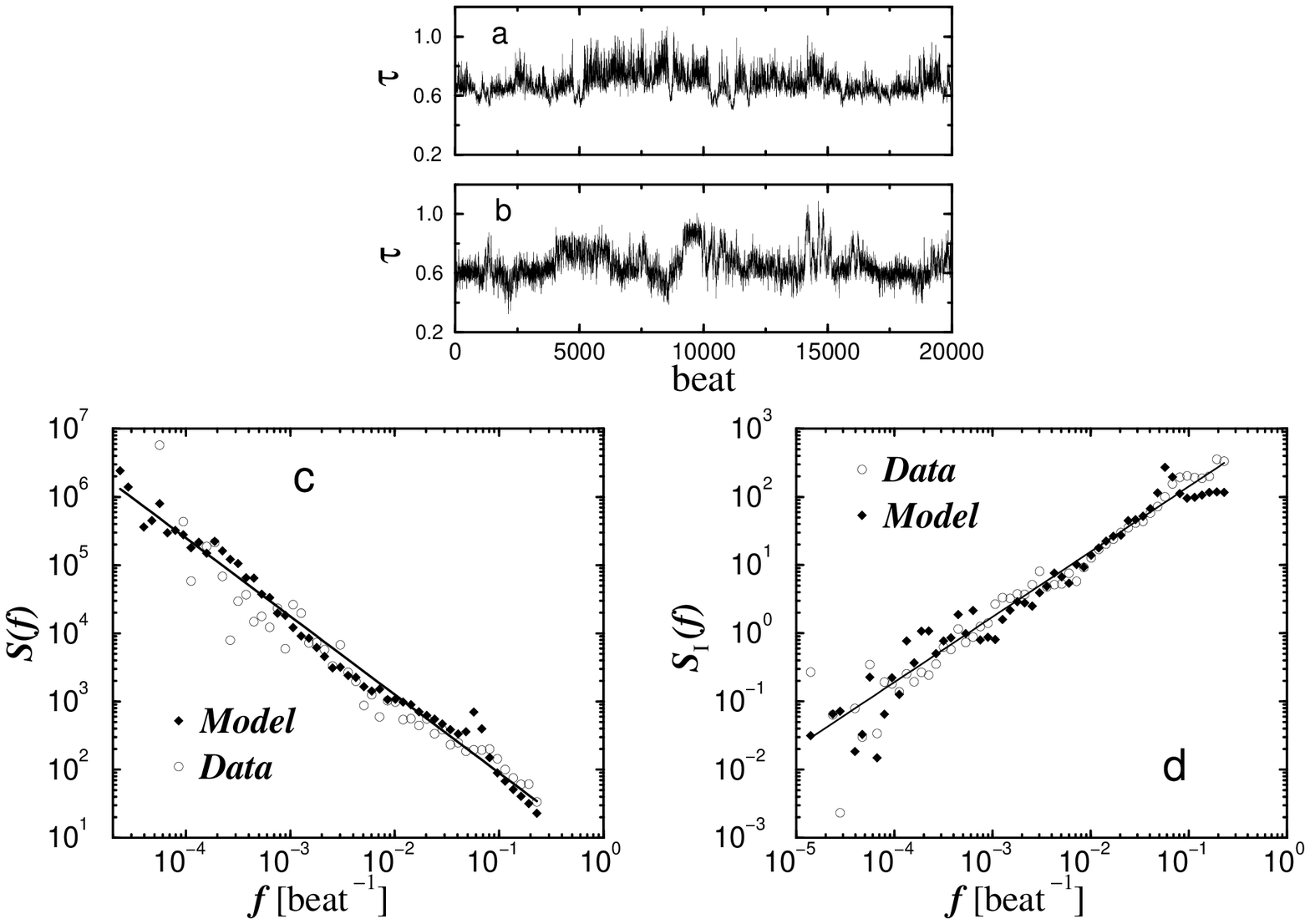}}}
\vspace*{-3.7cm}
\caption{ 
(a) Sequence of interbeat intervals $\tau$ for a healthy
individual. (b) Sequence of interbeat intervals for the model with
parameters $N=7$, $w_{SA} = w_{SS} = w_{PS} / 3 = 0.01$sec.  For the
results presented here, $T_i$ is randomly chosen from an exponential
distribution with average $T_{\rm lock} = 1000$ beats, and $\eta$ is a
symmetrical exponential distribution with zero average and standard
deviation $\sigma = 0.5$.  The preferred values of the interbeat
intervals for the different inputs were picked according to the
following rules: (1) $\tau_{SA} = 0.6$sec, (2) $\tau_{PS}$ are
randomly selected from an uniform distribution in the interval $[0.9,
1.5]$sec, and (3) the $\tau_{SS}^j$'s are randomly selected from an
uniform distribution in the interval $[0.2, 1.0]$sec.  We note that
the actual value of the preferred interbeat intervals of the different
inputs and the ratio between their weights are physiologically
justified and are of no significance for the dynamics --- they just
set the range for the fluctuations of $\tau$, chosen to correspond to
the empirical data. Also, any change of the shape of the distribution
of the noise term or of the locking times leaves the reported here
statistical properties of the generated signal unchanged.  (c) Power
spectra of the interbeat intervals $\tau(n)$ from the data and the
model described by the relation $S(f) \sim 1/f^{1.1}$.  The presence
of patches in both heart and model signals lead to observable
crossovers embedded on this $1/f$ behavior at different time scales.
The local exponent $\beta$ from the power spectrum of 24h records
($\approx 10^5$ beats) for $20$ healthy subjects shows a persistent
drift, so {\it no true scaling exists}.  (d) Power spectra of the
increments in $\tau(n)$.  The model and the data both scale as power
laws with exponents close to one. Since the non-stationarity is
reduced, crossovers are no longer present.  Here the local exponent
$\beta_I$ fluctuates around an average value close to one, so {\it
true scaling does exist}.
}
\label{f-results}
\end{figure}

\begin{figure}
\narrowtext
\vspace*{0.0cm}
\centerline{
\epsfysize=1.4\columnwidth{\epsfbox{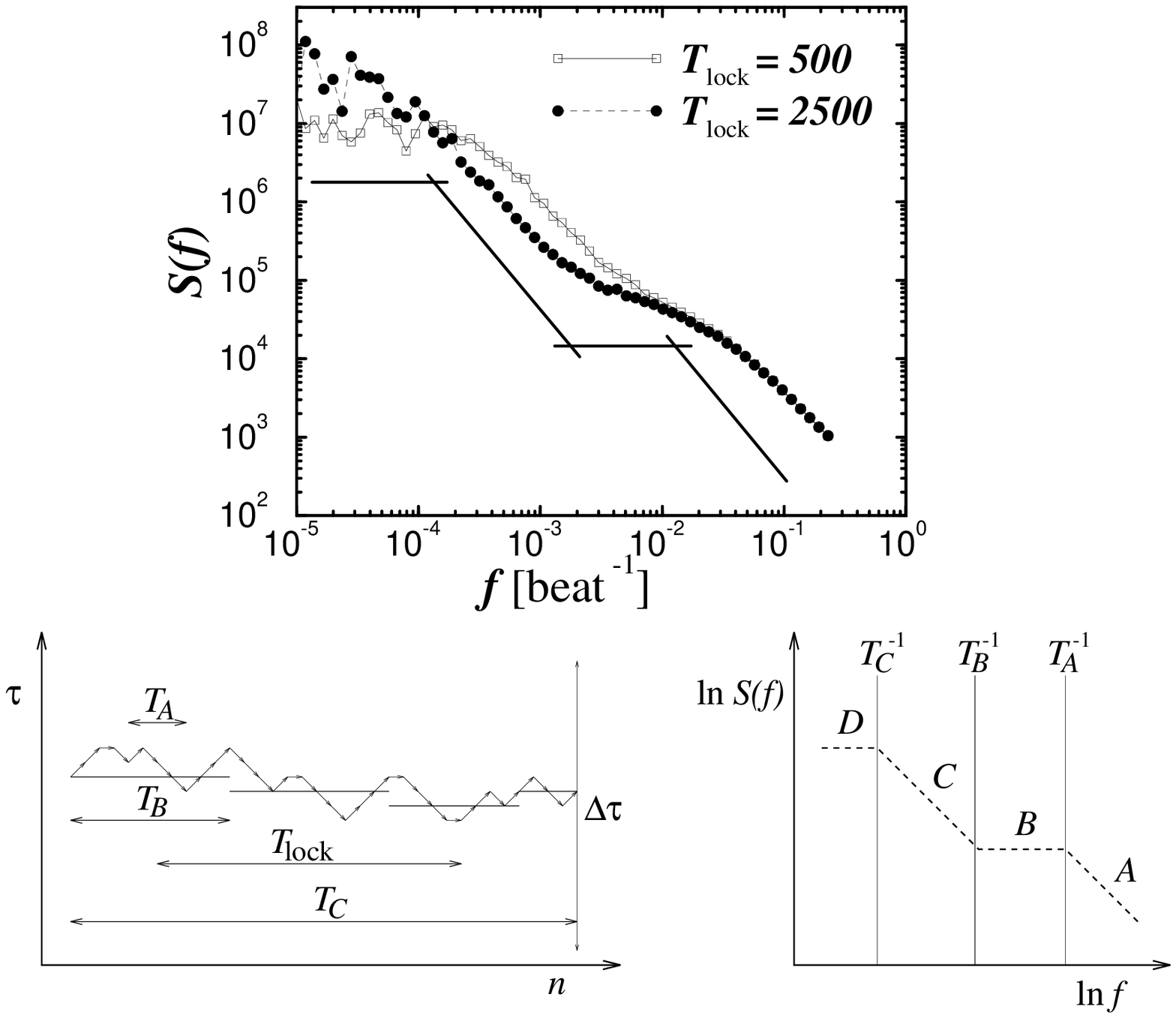}}}
\vspace*{-2.5cm}
\caption{Top: Effect of the correlation time $T_{\rm lock}$ on the
scaling of the power spectrum of $\tau(n)$ for a generated signal comprising
$10^{6}$ beats. With increasing $T_{\rm lock}$, the power spectrum
does not follow a single power law but actually crosses over from a
behavior of the type $1/f^2$ at very small time scales (or high
frequencies), to a behavior of the type $1/f^0$ for intermediate time
scales, followed by a new regime with $1/f^2$ for larger time scales.
At very large time scales, another regime appears with flat power
spectrum. Bottom: Schematic diagram illustrating the origin of the
different scaling regimes in the power spectrum of $\tau(n)$.  
For very short time scales, the noise
will dominate, leading to a simple random walk behavior and $1/f^2$
scaling (Region A).  For time scales longer than $T_A$, the
attraction towards the ``average preferred level'' of
all inputs will dominate, leading to a flat power spectrum (Region B,
see also Fig.~\protect\ref{f-walk}b).  However, after a time $T_B$ (of
the order of $T_{\rm lock} / N$), the preferred level of one of the
inputs will have changed, leading to the random drift of the average
preferred level and the consequent drift of the walker towards it.
So, at these time scales, the system can again be described as a
simple random walker and we expect a power spectrum of the type
$1/f^2$ (Region C).  Finally, for time scales larger than $T_C$, the
walker will start to feel the presence of the bounds on the
fluctuations of the preferred levels of the inputs.  Thus, the power
spectrum will again become flat (Region D).  
}
\label{f-slopes}
\end{figure}

  To {\it qualitatively} test the model, we first compare the time series
generated by the stochastic feedback model and the healthy heart
and find that both signals display complex
variability and patchiness (Fig.~\ref{f-results}a). 
To {\it quantitatively} test the model, we
compare the statistical properties of heart data with the predictions
of the model:

(a) We first test for long-range anti-correlations in the
interbeat intervals, which exist for healthy heart dynamics
\cite{Peng93}.  These anti-correlations can be uncovered by
calculating power spectra, and we see (Fig.~\ref{f-results}) that the
model correctly reproduces the observed long-range anti-correlations.
In particular, we note that the non-stationarity of both the data and
model signals leads to the existence of several distinct scaling
regimes in the power spectrum of $\tau(n)$ 
(Figs.~\ref{f-results}~and~\ref{f-slopes}). 
The stochastic feedback mechanism thus enables us 
to explain the formation of regions (patches) in the time series with
different characteristics (see caption to Fig.~\ref{f-slopes}).

(b) By studying the power spectrum of the increments we are able to
circumvent the effects of the non-stationarity.  Our results show that
$1/f$-scaling is indeed observed for the power spectrum of the
increments, both for the data and for the model (Fig.~\ref{f-results}).

\begin{figure}
\narrowtext
\vspace*{0.0cm}
\centerline{
\epsfysize=1.5\columnwidth{\epsfbox{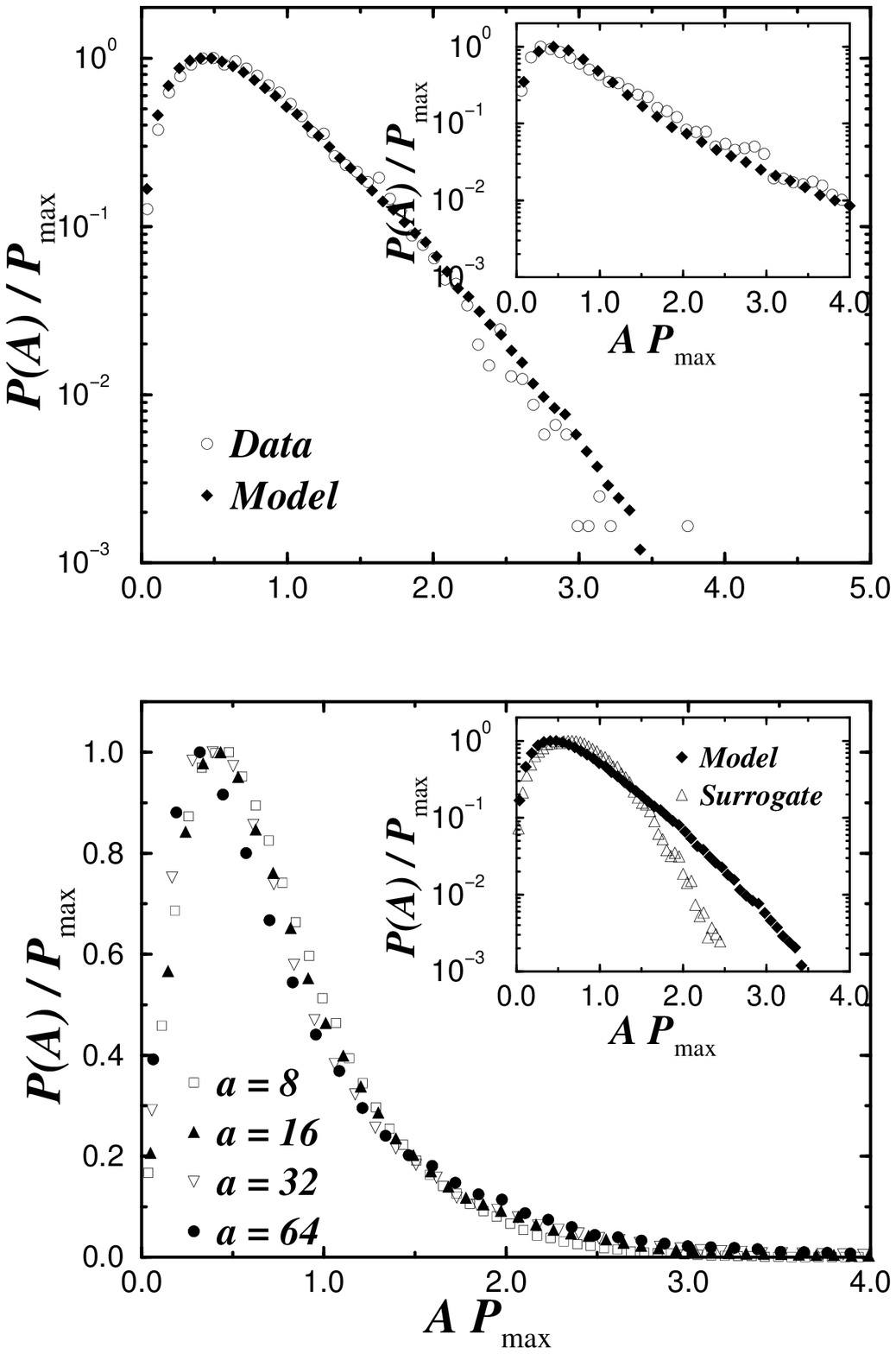}}}
\vspace*{-1.5cm}
\caption{ 
We apply to the signal generated by the model the wavelet transform
with fixed scale $a$, then use the Hilbert transform to calculate the
amplitude $A$. Top: normalized histogram $P(A)$ for the data (6h
daytime) and for the model (with the same parameter values as in
Fig.~\protect\ref{f-results}), and for $a=8$ beats ($\approx 40s$).
The inset shows a similar plot for data collected during the night and
for the model with $N < w_{PS} / w_{SS}$.  Note that the distribution
is broader for this case with large values for the amplitudes
deviating from the exponential tail. Bottom: we test the stability of
the analysis for the model at different time scales $a$. The
distribution is stable over a wide range of scales (identical to the
range observed for heart data) and indicates statistical
self-similarity in the variations at different time scales. We test
the generated signal for nonlinearity and Fourier phase correlations,
creating a surrogate signal by randomizing the Fourier phases of the
generated signal but preserving the power spectrum (thus, leaving the
results of Fig.~\protect\ref{f-results} unchanged).  The histogram of
the amplitudes of variations for the surrogate signal follows the
Rayleigh distribution, as expected theoretically (see inset). Thus the
observed distribution which is universal for healthy cardiac dynamics,
and reproduced by the model, reflects the Fourier phase
interactions. }
\label{f-wav}
\end{figure}

(c) We calculate the probability density $P(A)$ of the
amplitudes $A$ of the variations of interbeat intervals through the
wavelet transform.  It has been shown that the analysis of sequences
of interbeat intervals with the wavelet transform \cite{Grossmann85}
can reveal important scaling properties \cite{Muzy94} for the
distributions of the variations in complex nonstationary signals.  In
agreement with the results of Ref.~\cite{Ivanov96}, we find that the
distribution $P(A)$ of the amplitudes $A$ of interbeat interval
variations for the model decays exponentially---as is observed for
healthy heart dynamics (Fig.~\ref{f-wav}). We hypothesize that this
decay arises from nonlinear Fourier phase interactions and is related
to the underlying nonlinear dynamics.  To test this hypothesis, we
perform a parallel analysis on a surrogate time series obtained by
preserving the power spectrum but randomizing the Fourier phases of a
signal generated by the model (Fig.~\ref{f-wav}); $P(A)$ now follows
the Rayleigh distribution $P(A)\sim Ae^{-A^{2}}$ \cite{Stratonovich},
since there are no Fourier phase correlations.

(d) For the distribution displayed in Fig.~\ref{f-wav},
we test the stability of the scaling form at different time scales; we
find that $P(A)$ for the model displays a scaling form stable over a
range of time scales identical to the range for the data
(Fig.~\ref{f-wav}) \cite{Ivanov96}.  Such time scale invariance
indicates statistical self-similarity \cite{Bassingthwaighte94}.

A notable feature of the present model is that in addition to the
power spectra, it accounts for the functional form and scaling properties of
$P(A)$, which as we show are independent from the power spectra.

The model has a number of parameters, whose values may vary from one
individual to another, so we next study the sensitivity of our results
to variations in these parameters. We find that the model is robust to
parameter changes.  The value of $T_{\rm lock}$ and the strength of
the noise $\sigma$ are crucial to generate dynamics with scaling
properties similar to those found for empirical data. We find that the
model reproduces key features of the healthy heart dynamics for a wide
range of time scales ($500 \le T_{\mbox{\scriptsize lock}} \le 2000$)
and noise strengths ($0.4 \le \sigma \le 0.6$).  The model is
consistent with the existence of an extended ``zone'' in parameter
space where scaling behavior holds, and our picture is supported by
the variability in the parameters for healthy individuals for which
similar scaling properties are observed.

  The model, and the data which it fits, support a revised view of
homeostasis that takes into account the fact that healthy systems
under basal conditions, while being continuously driven away from
extreme values, do not settle down to a constant output.  Rather, a
more realistic picture may involve nonlinear stochastic feedback
mechanisms driving the system.

\vspace*{-2.0cm}

\vspace*{-3.cm}

\vspace*{-2.0cm}

\vspace*{-2.0cm}

\begin{references}


\bibitem{Bernard}
C.~Bernard, {\it Les Ph\'{e}nom\'{e}nes de la Vie} (Paris, 1878);
W.~B.~Cannon, Physiol. Rev. {\bf 9}, 399 (1929).

\bibitem{Akselrod81}
M.~F.~Shlesinger and B.~J.~West, in {\it Random Fluctuations and Pattern 
Growth} (Kluwer, Dordrecht, 1988).

\bibitem{Wax54}
G.~H.~Weiss, {\it Aspects and Applications of the Random Walk}
(Elsevier Science B.V., North-Holland, New-York, 1994).


\bibitem{Mackey77}
M.~G.~Rosenblum and J.~Kurths, Physica A {\bf 215}, 439 (1995);
H.~Seidel and H.~P.~Herzel, in {\it Modelling the Dynamics of 
Biological Systems\/} E.~Mosekilde and O.~G.~Mouritsen, Eds. 
(Springer-Verlag, Berlin, 1995);
J.~M.~Hausdorff and C.~-K.~Peng, Phys. Rev. E {\bf 54}, 2154 (1996).


\bibitem{Berne96}
R.~M.~Berne and M.~N.~Levy, {\it Cardiovascular Physiology\/} 6th ed.
(C.V. Mosby, St. Louis, 1996).

\bibitem{Peng93}
C.~-K.~Peng {\it et al.\/}, Phys. Rev. Lett. {\bf 70}, 1343 (1993).

\bibitem{Grossmann85}
I.~Daubechies, Comm. Pure and Appl. Math. {\bf 41}, 909 (1988).

\bibitem{Muzy94}
J.~F.~Muzy, E.~Bacry, and A.~Arneodo, Int. J. Bifurc. Chaos {\bf 4},
245 (1994);
A.~Arneodo~{\it et al.\/}, Physica D {\bf 96}, 291 (1996).

\bibitem{Ivanov96} 
P.~Ch.~Ivanov {\it et al.\/}, Nature {\bf 383}, 323 (1996).

\bibitem{Stratonovich}
R.~L.~Stratonovich, {\it Topics in the Theory of Random Noise\/}
(Gordon and Breach, New York, 1981).

\bibitem{Bassingthwaighte94}
J.~B.~Bassingthwaighte, L.~S.~Liebovitch, and B.~J.~West, {\it Fractal
Physiology} (Oxford Univ. Press, New York, 1994).

\end{references}
\end{document}